\documentclass[preprint,11pt]{elsarticle}

\makeatletter
\def\ps@pprintTitle{%
  \let\@oddhead\@empty
  \let\@evenhead\@empty
  \def\@oddfoot{\reset@font\hfil\thepage\hfil}
  \let\@evenfoot\@oddfoot
}
\makeatother

\usepackage[top=1.4in, bottom=1.4in, left=1.5in, right=1.5in]{geometry}
\usepackage{amssymb}
\usepackage{mathtools}
\usepackage{mathrsfs}
\usepackage{polski}
\usepackage[utf8]{inputenc}
\usepackage[french,polutonikogreek,polish]{babel}

\newcommand{\im}{{\rm i}}
\newcommand{\dee}{{\rm d}}
\journal{}

\begin{document}

\begin{frontmatter}



\title{Topological phases: isomorphism, homotopy and $K$-theory}


\author{Guo Chuan Thiang}

\address{School of Mathematical Sciences, The University of Adelaide, SA 5005, Australia\footnote{Present address}, \\ Mathematical Institute, University of Oxford, OX2 6GG, United Kingdom.}

\begin{abstract}
Equivalence classes of gapped Hamiltonians compatible with given symmetry constraints, such as those underlying topological insulators, can be defined in many ways. For the non-chiral classes modelled by vector bundles over Brillouin tori, physically relevant equivalences include isomorphism, homotopy, and $K$-theory, which are inequivalent but closely related. We discuss an important subtlety which arises in the chiral Class AIII systems, where the winding number invariant is shown to be  relative rather than absolute as is usually assumed. These issues are then analyzed and reconciled in the language of $K$-theory.
\end{abstract}


\end{frontmatter}


\section{Introduction}
\label{section:introduction}
The study of quantum phases of matter has benefited greatly from numerous insights and techniques from topology. A recent idea proposes that the homotopy groups of the stable classical groups \cite{bott1959stable} can equally well classify topological phases of free-fermions. Together with the symmetry class determined by the information of charge-conjugation and time reversal symmetries \cite{ryu2010topological,heinzner2005symmetry}, one is led to a Periodic Table of gapped topological phases \cite{kitaev2009periodic}, \emph{provided the $K$-theory groups in such a table are properly interpreted}. There are several important subtleties in the mathematics and physical interpretation of these $K$-theory groups, which appear to have been overlooked in the literature. Some of these are explained and rectified in \cite{thiang2014on}. Independently of the $K$-theoretic approach, any classification scheme must refer to a well-defined family of physical systems, as well as clearly stated equivalence relations defining the classes. 

A common meta-definition of a topological phase is one which cannot be smoothly or continuously deformed into a ``trivial phase'' while maintaining certain constraints, but this leaves open the question of where this deformation takes place in, and indeed, the precise object being deformed. Furthermore, algebraic operations on phases are not included in such a definition, whereas useful topological invariants usually have extra structure such as that of a group, a ring, or even functorial properties. For example, the symbol $\mathbb{Z}$ is often referred to as the classification object for topological phases in certain symmetry classes. Whether this means that the phases form an abelian group or merely a countably infinite set is an important question to address. Indeed, one should also check whether the algebraic operations, such as composition and inverses (where present), have meaningful counterparts in the physical interpretation.

In an ``absolute'' classification scheme, one tries to assign labels to the phases in an intrinsic way which does not depend on any physically irrelevant choices. In particular, the classification object (which is often an abelian group) should provide invariants of the \emph{isomorphism classes} of objects in an appropriate category modelling the physical phases. For example, it is usual to model the valence bands of (non-chiral) Class A band insulators using complex vector bundles over the Brillouin torus $\mathbb{T}^d$. Then, the appropriate category of objects comprises the isomorphism classes of such bundles. Similarly, the non-chiral Class AI (resp.\ Class AII) insulators can be modelled using Real (resp.\ Quaternionic) vector bundles, which come with a natural notion of isomorphism which respects the Real (resp.\ Quaternionic) structure coming from time-reversal symmetry \cite{kitaev2009periodic,ryu2010topological,freed2013twisted,thiang2014on,de2014classification2}. There are natural topological invariants which can distinguish between non-isomorphic (complex/Real/Quaternionic) vector bundles \cite{de2014classification2}. A prominent example in the two-dimensional case in Class A is the first Chern number, which has been linked to the quantized Hall conductivity, and more recently, the experimentally verified Chern insulator \cite{jotzu2014experimental}. The Chern number is an invariant of the isomorphism class of the valence vector bundle, or indeed, its virtual class in $K$-theory. It may also be construed as a homotopy invariant, as we explain in Section \ref{section:ungradedisomorphismhomotopy}.

On the other hand, there is also the general paradigm of a homotopy classification. It begins with the collection of possible Hamiltonians compatible with certain pre-specified constraints which typically arise from symmetry considerations and possibly a gapped condition. These Hamiltonians are presumed to form a topological space $Y$, in which two Hamiltonians are considered to be homotopically equivalent, if they can be connected by a continuous (or even smooth) path in $Y$. It is then natural to declare that the \emph{set} of (allowed) phases, up to homotopy, is the set of path-components $\pi_0(Y)$. However, it is not clear that isomorphic Hamiltonians in $Y$ (where ``isomorphism'' is assumed to be defined in some appropriate way which at least respects the symmetries) must also be homotopic, i.e.\ in the same path-component. Mathematically, this is due to the possibility that the automorphism group for an object in some category, when appropriately topologized, need not be connected. 

This presents a problem for the notion of an \emph{absolute} phase in a homotopy classification in the above sense. Suppose that $H_1, H_2\in Y$ are isomorphic but not through a homotopy, and that $H$ describing some other system is also isomorphic to $H_1,H_2$. Then there is an ambiguity in assigning the element of $\pi_0(Y)$ which corresponds to the ``absolute'' phase of $H$. Such an ambiguity arises for the chiral classes\footnote{It should be noted that the conventions for symmetry classes in the early literature \cite{heinzner2005symmetry}, as well as \cite{zirnbauer2014bott}, do not refer to chiral symmetries (or Hamiltonian-reversing symmetries) as proper symmetries.}, as explained in Section \ref{section:explicitmodel}. In fact, finding a  proper definition for the category of \emph{chiral vector bundles} is a very subtle issue, and has only been accomplished very recently \cite{de2015chiral}. It involves making certain choices of reference maps implementing the chiral symmetry, and only the \emph{relative} phase with respect to such choices has an absolute meaning. This inherent ambiguity motivates the central idea of this paper: a \emph{relative} classification scheme, in which topological obstructions between Hamiltonians, rather than the Hamiltonians themselves, are unambiguously classified.

\section{A closer look at the winding number invariant for Class AIII band insulators}
\label{section:explicitmodel}
\subsection{Class A band insulators}
We will model a band insulator in $d$-dimensions using a complex Hermitian vector bundle over the Brillouin torus $\mathbb{T}^d$ of unitary characters for the group $\mathbb{Z}^d$ of translational symmetries of an underlying crystal lattice. As an example, consider a tight-binding model in $d=1$ with Hilbert space $l^2(\mathbb{Z})$, where $\mathbb{Z}$ labels the atomic positions as a \emph{set} on which the translations act. Upon choosing an origin and orientation, we can identify $\mathbb{Z}$ as a group, and perform a Fourier transform $l^2(\mathbb{Z})\rightarrow L^2(\mathbb{T})$. Here, $\mathbb{T}\cong S^1$ is the Pontryagin dual of $\mathbb{Z}$ --- a unitary character labelled by $k\in\mathbb{R}\,(\mathrm{mod}\, 2\pi)$ takes $n\in\mathbb{Z}$ to $e^{\im n k}\in\mathrm{U}(1)$.

We can regard $L^2(\mathbb{T})$ as the square-integrable sections of a Hermitian line bundle $E\rightarrow\mathbb{T}$, with $E$ interpreted as a single valence band. We note that as a line bundle, the trivialization $E\cong\mathbb{T}\times\mathbb{C}$ can be realized in many ways. For example, the vector $\delta_0\in l^2(\mathbb{Z})$, which takes the value $1$ at $n=0$ and vanishes everywhere else, gets Fourier transformed into the constant function $q_0:k\mapsto 1, k\in\mathbb{T}$, which can be taken as the global non-zero section defining a trivialization of $E$. The other basis vectors $\delta_n, n\in\mathbb{Z}$ get transformed into the functions $q_n:k\mapsto e^{-\im nk}$ with respect to this trivialization. Note that each $q_n$ defines a global non-zero section of $E$, and thus provides another possible trivialization of $E$. 

Although we can nominally define a winding number invariant for each such choice of trivialization, the choice of origin for the original ``reference'' trivialization is arbitrary, and we do not consider such choices to have physical importance. These choices relate to the $\mathrm{U}(1)$ phase freedom in choosing the Bloch eigenstate at each point in the Brillouin zone. More generally, one is usually interested in gauge-invariant quantities for the vector bundle of filled valence states associated to the Fermi projection of some family of Bloch Hamiltonians over the Brillouin torus, but not in the winding numbers of its gauge transformations. We note that the \emph{geometrical} meaning of different choices of trivializations has recently been explored in \cite{fruchart2014parallel}.

\subsection{Class AIII band insulators, winding numbers, and odd Chern characters}\label{subsection:windingnumbers}
Gapped Hamiltonians in the symmetry class AIII are characterized by the presence of a sublattice (also called chiral) symmetry $S$, which is unitary, squares to the identity, and anticommutes with the Hamiltonian. As is usual in the literature, we adjust the energy scale so that $0$ lies in the energy gap, and regard a gapped Hamiltonian $H$ to be homotopic to its spectral flattening into a self-adjoint grading operator $\Gamma=\mathrm{sgn}(H)$. For the purposes of a homotopy classification, we need only concern ourselves with $\Gamma$. Then a simplified mathematical model of a Class AIII band insulator in $d$ spatial dimensions (with $\mathbb{Z}^d$ translational symmetry) is a $\mathbb{Z}_2$-graded complex hermitian vector bundle $E$ over the Brillouin torus $\mathbb{T}^d$, equipped with an \emph{odd} bundle automorphism $S$ commuting with the bundle projection. The $\mathbb{Z}_2$-grading corresponds to the spectrally-flattened Hamiltonian, and distinguishes the conduction bands from the valence bands. An inspection of various Periodic Tables in the literature \cite{kitaev2009periodic,ryu2010topological} shows that we should expect a $\mathbb{Z}$-valued topological invariant for Class AIII insulators in $d=1$. We recall the construction of this ``winding number invariant'' \cite{ryu2010topological}, before proceeding to explain its rather subtle relative nature.

\subsubsection*{A single valence band in $d=1$}
First consider two-band models, so $E$ is a rank-two bundle over $\mathbb{T}\cong S^1$. Since we require $S\Gamma=-\Gamma S$, we cannot have $S=\mathrm{id}$. The conduction and valence bands are each line bundles, as are the $+1$ and $-1$ eigenbundles of $S$. In order to write down concrete fibrewise matrices for $\Gamma$ and $S$, we \emph{choose} a reference trivialization $E\mapsto\mathbb{T}\times \mathbb{C}^2$ such that the fibrewise matrices $S(k)$ for $S$ are diagonal, i.e.\ $S(k)=\begin{pmatrix} 1 & 0 \\ 0 & -1 \end{pmatrix}$ for all $k\in [0,2\pi]$, $S(0)=S(2\pi)$. Note that the choice of such a trivialization is not unique: there is a $\mathrm{U}(1)\times\mathrm{U}(1)$ gauge freedom which does not affect the matrices $S(k)$.

A $\Gamma$ which is compatible with the action of $S$ on $E$ must have Bloch Hamiltonians $\Gamma(k)$ which are off-diagonal in the reference basis. Since $\Gamma(k)^2=1$ and $\Gamma(k)^\dagger=\Gamma(k)$, we must have 
\begin{equation}
	\Gamma(k)=\begin{pmatrix} 0 & q(k) \\ q(k)^\dagger & 0\end{pmatrix}
\end{equation}
for some continuous function $q:\mathbb{T}\rightarrow \mathrm{U}(1)\cong S^1$. A homotopy between two such functions $q$ and $q'$ corresponds exactly to a homotopy between the $\Gamma$ and $\Gamma'$ (within the space of operators\footnote{This can be made more precise in the language of $C^*$-algebras and projections, see Section \ref{section:ungradedisomorphismhomotopy}.}) which they determine. 

Therefore, the set of phases (compatible with the $S$), up to homotopy, is $\pi_1(\mathbb{T})\cong \mathbb{Z}$. It is natural to take as a reference Hamiltonian (the ``zero'' phase) the grading defined by the constant matrices $\Gamma_0(k)=\begin{pmatrix} 0 & 1 \\ 1 & 0\end{pmatrix}$. Thus $\Gamma_0$ corresponds to the constant function $q_0:k\mapsto e^{\im 0 k}=1$. As we explain below, this reference Hamiltonian is not invariantly defined and is implicitly dependent on the initial reference trivialization which diagonalizes $S$.

Each $n\in\mathbb{Z}=\pi_1(\mathbb{T})$ has a representative function $q_n(k)=e^{-\im nk}$ with winding number $n$. Having chosen the reference $\Gamma_0$, each label $n\in\mathbb{Z}$ can now be associated to the compatible Hamiltonian $\Gamma_n$ defined by $\Gamma_n(k)\coloneqq\begin{pmatrix} 0 & e^{-\im nk}\\ e^{\im nk} & 0 \end{pmatrix}$. Different $n$ correspond to non-homotopic $q_n$ and non-homotopic $\Gamma_n$. However, the $\Gamma_n$ are all ``isomorphic'' in the following sense. Define the bundle maps (or $\mathrm{U}(1)\times\mathrm{U}(1)$ gauge transformations) $U_n$ by the fibrewise operators $U_n(k)=\begin{pmatrix} q_n(k) & 0 \\ 0 & 1\end{pmatrix}$, which are unitary and commute with $S(k)$. Then
\begin{eqnarray}
U_n(k)\Gamma_0(k) U_n(k)^{-1}&=&\begin{pmatrix}e^{-ink} & 0 \\ 0 & 1\end{pmatrix}\begin{pmatrix} 0 & 1 \\ 1 & 0\end{pmatrix}\begin{pmatrix}e^{ink} & 0 \\ 0 & 1\end{pmatrix}\nonumber\\&=&\begin{pmatrix} 0 & e^{-ink} \\ e^{ink} & 0\end{pmatrix}=\Gamma_n(k).
\end{eqnarray}
More generally, we have $U_n\Gamma_mU_n^{-1}=\Gamma_{m+n}$. Thus the $\Gamma_n$ differ from each other only by a redefinition of the choices of eigenvectors for $S$. Put in a different way, what we call $\Gamma_n$ with respect to one initial choice of trivialization (diagonalizing $S$) would have been called $\Gamma_0$ in another trivialization.

We conclude that the winding number $n\in\mathbb{Z}$, which is a homotopy invariant of the \emph{map} $q_n$, should be interpreted as a label for the \emph{relative} obstruction (in a homotopy sense) between $\Gamma_n$ and a reference phase $\Gamma_0$. Indeed, these obstructions inherit the \emph{group} structure\footnote{Fortuitously $\mathrm{U}(1)\cong S^1$ is itself a group, so the group structure on the homotopy classes of (based) maps $[\mathbb{T},\mathrm{U}(1)]$ can be taken in two ways: (1) by concatenating loops, or (2) pointwise multiplication of loops. Both choices lead to the group $\mathbb{Z}$. The first choice is used for the fundamental group of any space, not necessarily a topological group like $\mathrm{U}(1)$. However, it is the second one which generalizes when we go beyond rank-2 models.} of $\pi_1(\mathbb{T})$. This relative point of view has \emph{absolute} meaning as it does not depend on the initial choice of trivialization defining the reference phase $\Gamma_0$. We also see that a proper notion of isomorphism classes of Class AIII band insulators is quite tricky to define --- it certainly should not simply be ``isomorphism of graded vector bundles with $S$-action'' --- see \cite{de2015chiral} for a detailed analysis.

\subsubsection*{AIII versus BDI}
The BDI symmetry class \cite{ryu2010topological} has an antilinear time-reversal symmetry $T$ which squares to the identity and satisfies $T\Gamma(k)T^{-1}=\Gamma(-k)$, along with a chiral symmetry $S$. Each of the $\Gamma_n$ described in the previous subsection is compatible with a time-reversal symmetry, namely the transformation which takes a vector over $k$ to its complex conjugate vector over $-k$. However, this ``accidental'' symmetry is not compatible with the other Hamiltonians within the same class of Hamiltonians (having the same winding number). Each $\Gamma_n$ is merely a convenient and highly symmetric choice of representative Hamiltonian in its AIII class. This is an important observation --- the correct symmetry class is determined by the symmetry constraints in question, not those of \emph{particular} compatible Hamiltonians. The latter may be compatible with some additional symmetries which are \emph{not} required to be preserved under the allowed deformations.

\subsubsection*{Winding number for $N$ valence bands in $d\geq1$}
We can also consider rank-$2N$ models for Class AIII insulators in $d\geq1$. Note that $S$ provides an isomorphism between the valence and conduction subbundles, which must both have rank $N$. There are homotopy invariants, generalizing the winding number introduced earlier, which can be associated to continuous maps $Q:\mathbb{T}^d\rightarrow \mathrm{U}(N)$ \cite{prodan2014non}. If we assume that the valence and conduction bands combine to form a trivial bundle\footnote{This is a simplifying assumption, and globally non-trivial $E$ may have further interesting features, see Section \ref{section:homotopichamiltonians}.} $E$, and that $S(k)=\mathrm{diag}(1_N,-1_N), k\in\mathbb{T}^d$ with respect to some trivialization $E\cong \mathbb{T}^d\times \mathbb{C}^{2N}$, then each $Q$ determines a flattened compatible Hamiltonian 
\begin{equation}
	\Gamma_Q(k)=\begin{pmatrix} 0 & Q(k) \\ Q(k)^\dagger & 0 \end{pmatrix}.\label{hamiltonianfromQ}
\end{equation}
This is the construction found in \cite{ryu2010topological,prodan2014non}. Non-homotopic $Q$ determine non-homotopic $\Gamma_Q$, and we may distinguish them by the homotopy classes of maps $[\mathbb{T}^d,\mathrm{U}(N)]$. Note that these classes can be given a group structure by pointwise multiplication in $\mathrm{U}(N)$.

Once again, there is a choice involved for the reference Hamiltonian $\Gamma_0(k)=\begin{pmatrix} 0 & 1_N \\ 1_N & 0 \end{pmatrix}$ associated to the constant map $k\mapsto 1_N$. This can be traced back to the $\mathrm{U}(N)\times\mathrm{U}(N)$ gauge freedom in the reference trivialization which diagonalizes $S$. Furthermore, two Hamiltonians $\Gamma_Q, \Gamma_{Q'}$ associated to two maps $Q, Q':\mathbb{T}^d\rightarrow \mathrm{U}(N)$, are related by $U_{Q'Q^\dagger}\Gamma_Q U_{Q'Q^\dagger}^{-1}=\Gamma_{Q'}$, where the unitary bundle map $U_{Q'Q^\dagger}$ is defined by
\begin{equation}
U_{Q'Q^\dagger}(k)=\begin{pmatrix} Q'(k)Q(k)^\dagger & 0 \\ 0 & 1_N \end{pmatrix},
\end{equation}
and respects the $S$-action.

Recall that in the $d=1=N$ example, we had $[\mathbb{T},\mathrm{U}(1)]\cong\pi_1(\mathbb{T})\cong \mathbb{Z}$, and the explicit computation of the winding number for $q_n:\mathbb{T}\rightarrow\mathrm{U}(1)$ is
\begin{equation}
\frac{\im}{2\pi}\int_0^{2\pi} q_n(k)^{-1}\,\dee q_n(k)=\frac{\im}{2\pi}\int_0^{2\pi}e^{\im nk}(-\im n)e^{-\im nk}\, \dee k=n.\label{windingnumber}
\end{equation}
If we regard the smooth map $q_n:S^1\rightarrow \mathrm{U}(1)$ as a unitary element of $C^\infty(S^1)$, the differential $1$-from in the integrand in \eqref{windingnumber} is, up to a constant factor, the \emph{odd Chern character} of $q_n$ (see \cite{park2008complex} for explicit formulae and generalizations). The odd Chern character is perhaps better understood in terms of $K$-theory (see Section \ref{section:ungradedisomorphismhomotopy} for some basic $K$-theory definitions). Every element of $K^{-1}(\mathbb{T})\cong \mathbb{Z}$ can be represented by a smooth unitary $Q:\mathbb{T}\rightarrow \mathrm{U}(N)$ in the matrix algebra $M_N(C^\infty(\mathbb{T}))$ for some $N\geq 1$. The Chern character of such a $Q$ is $\mathrm{tr}(Q^{-1}\dee Q)=\dee \log \det  Q$, which integrates over $\mathbb{T}$ to give $-2\pi\im$ times the winding number of $\det(Q):\mathbb{T}\rightarrow\mathrm{U}(1)$. To summarize: the winding number of $\det(Q)$ characterizes a topological obstruction between $\Gamma_Q$ and the reference $\Gamma_0$.

\subsubsection*{Odd Chern character in $d>1$}
More generally, $\mathbb{T}$ can be replaced by a higher dimensional compact manifold $X$ such as a higher dimensional Brillouin torus $\mathbb{T}^d$. The odd Chern character map takes a unitary $Q\in M_N(C^\infty(X))$ into a class in $H^\mathrm{odd}_\mathrm{de\,Rham}(X)$, which generally comprises higher odd-dimensional forms. The Chern character is insensitive to the (smooth) homotopy class of $Q$ and is even a homomorphism from $K^{-1}(X)$ to $H^\mathrm{odd}_\mathrm{de\,Rham}(X)$ \cite{park2008complex}. In fact, in our basic $d=1=N$ example, the (class of) the unitary $q_1$ (or $q_{-1}$) is actually the Bott generator for $K_1(C(S^1))\cong K^0(\mathbb{R}^2)\cong \mathbb{Z}$ (see Chapter 3.7 of \cite{gracia2001elements} for details), and the Chern character maps $q_1$ to $\dee k$. Actually, it is not necessary to define the Chern character in de Rham cohomology; indeed, an application of the \emph{noncommutative} odd Chern character can be found in \cite{prodan2014non2}.

\section{Connection between isomorphism and homotopy for non-chiral classes}
\label{section:ungradedisomorphismhomotopy}
Class A band insulators in $d$ dimensions can be classified by $K^0(\mathbb{T}^d)$, where a formal difference $E\ominus F$ of bundles has at least two different physical interpretations. On the one hand, $E$ can be regarded as the conduction bands and $F$ as the filled valence bands. On the other hand, $E\ominus F$ may simply be regarded as a formal difference between two valence bands $E$ and $F$, with the data of the conduction band deemed to be irrelevant. Both interpretations are also possible for the other two non-chiral classes, upon replacing the complex $K$-theory groups by their Real $KR$-theory (Class AI) or Quaternionic $KQ$-theory (Class AII) counterparts.

There are a number of features of the complex $K^0$ functor which differs from $K^{-1}$ (and similarly for the Real and Quaternionic $K$-theory functors). Most important for our purposes is the availability of pictures of $K^0(X)$ in terms of both isomorphism and homotopy classes of vector bundles over $X$, where the latter needs to be defined carefully. In particular, we are certainly not interested in homotopies of bundles as topological spaces, since they always contract onto their base space (which is fixed as $\mathbb{T}^d$). The relevant notion of homotopy is that of projections in the stabilised algebra $M_\infty(C(\mathbb{T}^d))$, as we explain below. The $K^{-1}$ functor is more about automorphisms of bundles and homotopies of such automorphisms, although it can be linked to $K^0$ by taking suspensions (which changes the base space) or through Clifford algebras and Karoubi triples \cite{karoubi1978k}.

\subsubsection*{The $K^0$ functor in terms of projections}
Complex vector bundles over a compact Hausdorff space $X$ correspond, by the Serre--Swan theorem, to (left) finitely-generated projective (f.g.p.) modules for the $C^*$-algebra $C(X)$ of continuous functions $X\rightarrow\mathbb{C}$. The latter are of the form $(C(X)^N)p$ for some projection $p$ in some matrix algebra $M_N(C(X))$. Here, $C(X)^N$ is the free $C(X)$-module of continuous sections of the trivial rank-$N$ bundle over $X$. Note that $p$ may also be considered to be the projection $p\oplus 0_{N'-N}\in M_{N'}(C(X))$ for any $N'\geq N$. For a unital $C^*$-algebra $\mathcal{A}$ (e.g.\ $\mathcal{A}=M_N(C(X))$ for a fixed $N$), there are a number of equivalence relations which may be imposed on its projections. There is \emph{unitary equivalence}, where $p_0\sim p_1$ if there exists a unitary $u\in\mathcal{A}$ such that $up_0u^{-1}=p_1$. There is also \emph{homotopy equivalence}, where $p_0\sim_hp_1$ if there is a norm-continuous path of projections in $\mathcal{A}$ from $p_0$ to $p_1$.

While $p_0\sim_hp_1$ implies $p_0\sim p_1$, the converse is not generally true \cite{blackadar1998k}. Nevertheless, the converse does hold if we regard the $p_i$ as projections in $M_2(\mathcal{A})$. To see this, we first note that for any two unitaries $u,v\in\mathcal{A}$, we can construct a path of unitaries $U_t$ in $M_2(\mathcal{A})$ between $U_0=\mathrm{diag}(uv,1)$ and $U_1=\mathrm{diag}(u,v)$, via
\begin{equation}
	U_t=\begin{pmatrix}u & 0 \\ 0 & 1\end{pmatrix}\begin{pmatrix}\cos \frac{\pi}{2}t & -\sin \frac{\pi}{2}t \\ \sin \frac{\pi}{2}t & \cos \frac{\pi}{2}t\end{pmatrix}\begin{pmatrix}v & 0 \\ 0 & 1 \end{pmatrix}\begin{pmatrix}\cos \frac{\pi}{2}t & \sin \frac{\pi}{2}t \\ -\sin \frac{\pi}{2}t & \cos \frac{\pi}{2}t\end{pmatrix},\quad t\in[0,1].\label{pathofunitaries}
\end{equation}
Taking $v=u^{-1}$, we see that there is a homotopy in $M_2(\mathcal{A})$ between $U_0=1$ and $U_1=\mathrm{diag}(u,u^{-1})$. If $p_0\sim p_1$ is implemented through $u$, then $P_t\coloneqq U_t\mathrm{diag}(p_0,0)U_t^{-1}$ is a homotopy in $M_2(\mathcal{A})$ between $P_0=\mathrm{diag}(p_0,0)$ and $P_1=\mathrm{diag}(p_1,0)$. 

When posing the question of whether two vector bundles over $X$ are ``homotopic'', a fixed background bundle for which the two bundles are subbundles is implicitly fixed. Typically, this ambient bundle is taken to be a trivial bundle $X\times\mathbb{C}^N$, and the homotopy in question is between the projections in $M_N(C(X))$ corresponding to the two subbundles. In this sense, homotopy and isomorphism are not necessarily equivalent. Nevertheless, the constructions in the previous paragraph show that isomorphic bundles can always be considered to be homotopic when placed within a larger ambient bundle. Indeed $K^0(X)\cong K_0(C(X))$ can be defined as the the Grothendieck group of the monoid of equivalence classes of projections in $M_\infty(C(X))=\varinjlim M_N(C(X))$, where either unitary equivalence or homotopy equivalence may be used (see Chapter 5 of \cite{blackadar1998k}).

\subsection{A relative view of $K^0(X)$}\label{section:relativek0}
There is an alternative picture of $K^0(X)$ due to Karoubi \cite{karoubi1978k}, which makes this idea of homotopy within an ambient bundle more explicit (also see \cite{nicolaescu1997generalized} for a less abstract presentation). The detailed construction can be found in the references, so we simply illustrate it with the simplest example of $X=\{\mathrm{pt}\}$. We have $K^0(\mathrm{pt})\cong\mathbb{Z}$ generated by the vector space $\mathbb{C}$. In Karoubi's picture, this generator corresponds to the class of the triple $(\mathbb{C},1,-1)$ which represents the (ordered) difference between the purely even grading $+1$ on $\mathbb{C}$ and the purely odd grading $-1$. Note that the ranks of the $-1$ eigenspaces of the grading operators differ by one. $K^0(\mathrm{pt})$ is generated by general triples of the form $(\mathbb{C}^N,\Gamma_1,\Gamma_2)$, representing the difference between grading operators $\Gamma_1, \Gamma_2$ on $\mathbb{C}^N$. Triples may be added by the direct sum operation on the vector spaces and the grading operators. For the definition of equivalence classes of triples, we allow each $\Gamma_i$ to be replaced by a grading operator homotopic to it, so we may write each class in the form \mbox{$[\mathbb{C}^N,1_{N-n}\oplus-1_n,1_{N-n'}\oplus-1_{n'}]$} for some $0\leq n,n'\leq N$. Also, a triple with homotopic $\Gamma_i$ is considered to be trivial.

Suppose (without loss of generality) that $n'\geq n\geq 0$, then the algebraic rules for triples summarized above allow us to write
\begin{eqnarray}
	[\mathbb{C}^N,\Gamma_1,\Gamma_2]&=&[\mathbb{C}^N,1_{N-n}\oplus-1_n,1_{N-n'}\oplus-1_{n'}]\nonumber\\
	&=&[\mathbb{C}^{N+n-n'},1_{N-n'}\oplus-1_n,1_{N-n'}\oplus-1_n]+[\mathbb{C}^{n'-n},1_{n'-n},-1_{n'-n}]\nonumber\\ &=&[\mathbb{C}^{n'-n},1_{n'-n},-1_{n'-n}]=(n'-n)[\mathbb{C},1,-1],	
\end{eqnarray}
and we associate $[\mathbb{C}^N,\Gamma_1,\Gamma_2]$ to the integer $n'-n$. In other words, the ``difference class'' of the triple $(\mathbb{C}^N,\Gamma_1,\Gamma_2)$ counts the \emph{change} in the rank of the $-1$ eigenspaces modulo homotopy and addition of trivial differences.

For a general compact Hausdorff space $X$, the triples generating the group $K^0(X)$ are of the form $(E,\Gamma_1,\Gamma_2)$, with $E$ a vector bundle over $X$ and $\Gamma_i$ gradings on $E$. We may assume $E$ to be trivial by augmenting $(E,\Gamma_1,\Gamma_2)$ by a trivial triple $(E^\perp,1,1)$, where $E\oplus E^\perp\cong X\times \mathbb{C}^N$. The ``difference class'' $[E,\Gamma_1,\Gamma_2]$ in $K^0(X)$ can be understood as labelling the \emph{change} in the  $-1$ eigenbundles when passing from $\Gamma_1$ to $\Gamma_2$, modulo homotopy and addition of trivial differences. Amongst others, this could include the change in rank and Chern classes of the $-1$ eigenbundles. A virtual class $[E\ominus F]$ in a more usual Grothendieck group definition of $K^0(X)$ corresponds to the difference-class of the triple $[E\oplus F, 1_E\oplus-1_F, -1_E\oplus 1_F]$.

An important advantage of the picture of $K^0$ in terms of Karoubi triples, is that it generalizes to the other symmetry classes while retaining the physical interpretation. In particular, the higher degree $K$-theory groups admit such a description. We give a brief account of this in Section \ref{section:homotopichamiltonians}.

\subsection{Aside: Homotopies of classifying maps}
There is a classifying space $\mathrm{BU}(N)$ for complex rank-$N$ bundles, which may be realised as the Grassmannian of $N$-planes in an infinite-dimensional Hilbert space. Every rank-$N$ bundle over $X$ can be realised as the pullback of the universal tautological bundle under some map $f:X\rightarrow \mathrm{BU}(N)$, and homotopic maps yield isomorphic bundles. For example, when $N=1$, the classifying space for complex line bundles is $\mathbb{C}\mathbb{P}^\infty$. In fact, $[X,\mathbb{C}\mathbb{P}^\infty]\cong H^2(X)$ where the right hand side is the ordinary (second) cohomology group of $X$ where the first Chern classes live.

Imposing symmetry constraints means that only bundles with some extra structure are allowed. One can imagine that a suitable classifying space exists for such bundles, and this is the point of view taken up in \cite{zirnbauer2014bott}. There, the relation of homotopy is imposed on the space of ``classifying maps'' which determine subbundles of an ambient trivial bundle of fixed rank $2N$, with the bundles compatible with the symmetry constraints. This can be understood in terms of projections in $M_{2N}(C(X))$ which are required to satisfy some additional conditions. As before, the relations of isomorphism (of bundles) and homotopy (of classifying maps) are not the same\footnote{The authors provide an example of this, see Example 3.1 of their paper.}, and their homotopy classification is finer than the isomorphism classification of subbundles. It should be noted that the authors require symmetries to commute with the Hamiltonian, so their conventions differ from those of other authors. In particular, the $S$-symmetry in Class AIII systems are regarded as pseudo-symmetries.

\section{Homotopic Hamiltonians and $K$-theory}
\label{section:homotopichamiltonians}
\subsection{A relative view of $K^{-1}(\mathbb{T}^d)$}
In Section \ref{section:explicitmodel}, we explained how the labelling of topological phases up to homotopy can be ambiguous when taken in an absolute sense. The ``primitive'' topological invariant there is the homotopy class of the \emph{map} $Q$ from the base space to a unitary group. Alternatively, $Q$ determines a class in a $K^{-1}$-group, and is detected by integrating the Chern character of $Q$ over the base space, yielding a numerical winding number invariant.

Let us rephrase our analysis in algebraic language. The sections of the ambient bundle $E$ over the Brillouin torus $\mathbb{T}^d$ form a free $C(\mathbb{T}^d)$-module $W=(C(\mathbb{T}))^{2N}$, and the bundle map $S$ translates into a module map $\mathsf{S}$ on $W$. We can think of $W$ as an ungraded module for the graded\footnote{All gradings in this paper are $\mathbb{Z}_2$-gradings. The symbol $\hat{\otimes}$ denotes the graded tensor product.} algebra $C(\mathbb{T}^d)\hat{\otimes}\mathbb{C}l_1$, where $\mathbb{C}l_1$ is the complex Clifford algebra whose odd generator is represented on $W$ by $\mathsf{S}$. Then the compatible flattened Hamiltonians on $W$ are precisely grading operators which turn $W$ into a \emph{graded} $C(\mathbb{T}^d)\hat{\otimes}\mathbb{C}l_1$-module, i.e.\ they are $S$-compatible Hamiltonians. As explicit examples for $d=1=N$, the $\Gamma_n$ associated with the maps $q_n:k\mapsto e^{-\im nk}$ are $S$-compatible Hamiltonians which are non-homotopic for different $n$. Furthermore, conjugating with $U_n=\mathrm{diag}(q_n,1)$ takes $\Gamma_m$ to $\Gamma_{m+n}$ for each $m$. More generally, the $\Gamma_Q$ associated with maps $Q:\mathbb{T}^d\rightarrow\mathrm{U}(N)$ are non-homotopic for non-homotopic $Q$. Conjugation by $U_{Q'}$ defined by $U_{Q'}(k)=\begin{pmatrix} Q'(k) & 0 \\ 0 & 1_N \end{pmatrix}$ takes $\Gamma_{Q}$ to $\Gamma_{Q'Q}$ for any $Q$. Furthermore, conjugation by $U_{Q'}$ followed by $U_{Q''}$ is the same as conjugation by $U_{Q''Q'}$.

We abstract these properties in terms of formal triples $(W,\Gamma,\Gamma')$ representing the obstruction in passing from $\Gamma$ to $\Gamma'$ within the space of compatible grading operators on $W$. Replacing $\Gamma$ or $\Gamma'$ by homotopic compatible grading operators should not change the class $[W,\Gamma,\Gamma']$ of the triple. The zero element is $[W,\Gamma_{Q_0},\Gamma_{Q_0}]$, where \emph{any} $Q_0$ may be used. The map $Q$ should be associated with the difference class $[W,\Gamma_{Q_0},\Gamma_{QQ_0}]$, where again any $\Gamma_{Q_0}$ may be used. The group operation on $[\mathbb{T}^d,\mathrm{U}(N)]$, namely $(Q',Q)\mapsto Q'Q$ (on representative functions), translates into the rule for triples
\begin{equation}
	[W,\Gamma_{Q_0},\Gamma_{QQ_0}]+[W,\Gamma_{QQ_0},\Gamma_{Q'QQ_0}]=[W,\Gamma_{Q_0},\Gamma_{Q'QQ_0}],\label{triplepathindependence}
\end{equation}
while the inverse map $Q\mapsto Q^\dagger$ becomes the rule
\begin{equation}
	-[W,\Gamma_{Q_0},\Gamma_{QQ_0}]=[W,\Gamma_{Q_0},\Gamma_{Q^\dagger Q_0}]=[W,\Gamma_{QQ_0},\Gamma_{QQ^\dagger Q_0}]=[W,\Gamma_{QQ_0},\Gamma_{Q_0}],\label{tripleinverse}
\end{equation}
which is just the obstruction taken in the opposite order. Note that \eqref{pathofunitaries} says that $[Q\oplus Q']=[QQ'\oplus 1]=[Q'Q]$. Thus, upon augmentation by trivial triples, we can compose triples by multiplying the unitaries associated to the triples as in \eqref{triplepathindependence}, or by direct sum.

What we have just described is precisely a model for $K_1(C(\mathbb{T}^d))\cong K^{-1}(\mathbb{T}^d)$ using Karoubi's triples \cite{karoubi1978k}, and adapted for the classification of obstructions between topological phases in \cite{thiang2014on}. It is the Class AIII version of our presentation of $K^0(X)\cong K_0(C(X))$ for Class A systems in Section \ref{section:relativek0} using similar triples. The classes of topological obstructions between compatible Class AIII insulators in dimension $d$ (with $N$ unrestricted) are thus given by the group $K^{-1}(\mathbb{T}^d)$, which in the special case $d=1$ is isomorphic to $\mathbb{Z}$.

\subsection{A relative view of $K_1(\mathcal{A})$}
For a general ungraded unital $C^*$-algebra $\mathcal{A}$, the group $K_1(\mathcal{A})$ may be defined as the abelian group generated by homotopy classes of unitaries in the matrix algebras $M_N(\mathcal{A}), N\geq 1$, with $[1]=0$ and $[u]+[v]=[u\oplus v]$ \cite{higson2000analytic}. Recall that $[u\oplus v]=[uv\oplus 1]=[uv]$, so composition of classes in $K_1$ may be realised on representative unitaries in a number of ways. Unitaries in 
$M_N(\mathcal{A})$ may also be interpreted as obstructions between Type AIII gapped Hamiltonians as follows. An ungraded f.g.p.\ $\mathcal{A}\hat{\otimes}\mathbb{C}l_1$-module $W$ may be written as $W_+\oplus W_-$, where the f.g.p.\ $\mathcal{A}$-modules $W_\pm$ are the $\pm 1$ eigenspaces of the operator $\mathsf{S}$ representing the Clifford generator. We may assume that $W_\pm=\mathcal{A}^Np_\pm$ for some projections $p_\pm$. If there is any compatible grading operator at all, it must be of the form $\Gamma=R_{+-}\oplus R_{-+}$ for $\mathcal{A}$-module maps $R_{+-}:W_+\rightarrow W_-$ and $R_{-+}:W_-\rightarrow W_+$. Since $\Gamma$ is self-adjoint and involutary, it follows that $R_{+-}$ is unitary and $R_{-+}$ is its adjoint map. Given any unitary $Q\in M_N(\mathcal{A})$, we can construct another compatible grading operator $\Gamma_Q\coloneqq R_{+-}Q^{-1}\oplus QR_{-+}$. 

Thus, the original grading $\Gamma$ plays the role of a reference grading operator, and non-homotopic unitaries $Q\in M_N(C(X))$ lead to non-homotopic $\Gamma_Q$. This construction generalises \eqref{hamiltonianfromQ}, which is the special case where $\mathcal{A}=C(\mathbb{T}^d), W=C(\mathbb{T}^d)^{2N}$ and $\mathsf{S}=\mathrm{diag}(1_{C(\mathbb{T}^d)^N},-1_{C(\mathbb{T}^d)^N})$. There are corresponding triples $(W,\Gamma_Q,\Gamma_{Q'})$ representing the obstruction between $\Gamma_Q$ and $\Gamma_{Q'}$, and such triples generate $K_1(\mathcal{A})$ in Karoubi's model. Note that in the more general setting, the $\mathcal{A}$-modules $W_\pm$ need not be free (corresponding to trivial vector bundles if $\mathcal{A}\cong C(X)$), so we can actually also define invariants belonging to $K_0(W_\pm)$ for a given triple $(W,\Gamma_Q,\Gamma_{Q'})$, see \cite{de2015chiral} for a detailed discussion.

\subsection{A unified picture of homotopic Hamiltonians using difference-groups}
The classes A and AIII are the so-called complex symmetry classes and do not have antiunitary symmetry constraints. In the presence of antiunitary charge-conjugation or time-reversal symmetries, there are eight real symmetry classes, each of which is associated to a Morita class of real Clifford algebras \cite{kitaev2009periodic,ryu2010topological,freed2013twisted,thiang2014on}. Compatibility of a grading operator (gapped Hamiltonian) entails graded commutation with the Clifford algebra action. More generally, the symmetry constraints determine a \emph{graded} symmetry $C^*$-algebra $\mathcal{A}$ \cite{thiang2014on}, which is real if there is at least one antiunitarily implemented symmetry element. An ungraded $\mathcal{A}$-module $W$, can be understood as an ambient representation space hosting the symmetries encoded by $\mathcal{A}$. There is a (possibly empty) set $\mathrm{Grad}_\mathcal{A}(W)$ of compatible grading operators turning $W$ into a graded $\mathcal{A}$-module. Such grading operators are precisely the spectrally-flattened symmetry-compatible gapped Hamiltonians. Furthermore, homotopic compatible Hamiltonians are precisely those whose grading operators are homotopic\footnote{In more detail, $\mathcal{A}$ is a $C^*$-algebra, and $W$ can be made a Banach space. Then the space of $\mathcal{A}$-module maps on $W$, which contains the compatible grading operators, can be topologized naturally.} in $\mathrm{Grad}_\mathcal{A}(W)$. 

For a graded $C^*$-algebra $\mathcal{A}$ (such as the symmetry algebra), the $K$-theoretic \emph{difference group} $\mathbf{K}_0(\mathcal{A})$ as defined in \cite{thiang2014on} is generated by triples $[W,\Gamma_1,\Gamma_2]$ where $W$ is an ungraded f.g.p.\ $\mathcal{A}$-module and $\Gamma_i\in\mathrm{Grad}_\mathcal{A}(W)$. Such a triple represents the obstruction in passing from $\Gamma_1$ to $\Gamma_2$ in a homotopic manner. The difference-group has a uniform interpretation which works not only for all ten standard symmetry classes, but also for more general symmetry algebras. Triples may be added by taking direct sums, and they satisfy the properties of path-independence and existence of inverses as in \eqref{triplepathindependence} and \eqref{tripleinverse}. In special cases, these difference groups are isomorphic to ordinary $K$-theory groups. For instance, we sketched a model of $K^{-1}(\mathbb{T}^d)\cong K_1(C(\mathbb{T}^d))$ using such triples at the beginning of this section.

\section*{Conclusion}
To summarize, we propose that $K$-theory be used as a way to obtain groups of topological obstructions between gapped Hamiltonians. It allows us to measure one phase \emph{relative} to another, and is suited to analyzing interfaces between two phases. This is in contrast with the idea of a topological classification of gapped Hamiltonians in an absolute sense up to homotopy, which can be problematic. Furthermore, ``absolute'' phases are really special cases of the relative picture in which a canonical zero phase (``vacuum'') can be defined. As we have seen in the Class AIII examples, such a phase need not be canonically available. We stress that the relative viewpoint is not completely new and certainly not controversial; it was mentioned in Kitaev's seminal work \cite{kitaev2009periodic}, and related notions of \emph{relative index} and \emph{charge deficiency} had previously been defined and applied to the Integer Quantum Hall Effect in \cite{avron1990quantum,avron1994charge}.

\section*{Acknowledgements}
The author acknowledges helpful discussions with Giuseppe de Nittis, Dan Freed, Emil Prodan, Keith Hannabuss, Andr\'{e} Henriques, Thomas Wasserman, and Martin Zirnbauer. He was funded by Balliol College, the Clarendon Fund, and the Australian Research Council Discovery Project grant DP110100072.


\begin{thebibliography}{0}
\bibitem{bott1959stable}
R. Bott, The stable homotopy of the classical groups, {\it Ann. of Math.} {\bf 70} (1959), 313--337.

\bibitem{ryu2010topological}
S. Ryu, A. P. Schnyder, A. Furusaki, and A. W. W. Ludwig, Topological insulators and superconductors: tenfold way and dimensional hierarchy, {\it New J. Phys.} {\bf 12} (2010) 065010.

\bibitem{heinzner2005symmetry}
P. Heinzner, A. Huckleberry, and M. R. Zirnbauer, Symmetry classes of disordered fermions, {\it Commun. Math. Phys.} {\bf 257} (2005) 725--771.

\bibitem{kitaev2009periodic}
A. Kitaev, Periodic table for topological insulators and superconductors, in {\it AIP Conf. Ser.} {\bf 1134} (2009) 22--30.

\bibitem{thiang2014on}
G. C. Thiang, On the ${K}$-theoretic classification of topological phases of matter, math-ph/1406.7366.

\bibitem{freed2013twisted}
D. S. Freed and G. W. Moore, Twisted equivariant matter, {\it Ann. Henri Poincar{\'e}} {\bf 14} (2013), 1927--2023.

\bibitem{de2014classification2}
G. de Nittis and K. Gomi, Classification of ``{Q}uaternionic'' {B}loch-bundles: {T}opological Insulators of type {AII}, math-ph/1404.5804.

\bibitem{jotzu2014experimental}
G. Jotzu, M. Messer, R. Desbuquois, M. Lebrat, T. Uehlinger, D. Greif, and T. Esslinger, Experimental realization of the topological {H}aldane model with ultracold fermions, {\it Nature} {\bf 515} (2014) 237--240.

\bibitem{zirnbauer2014bott}
M. R. Zirnbauer and R. Kennedy, Bott periodicity for $\mathbb{Z}_2$ symmetric ground states of gapped free-fermion systems, math-ph/1409.2537.

\bibitem{de2015chiral}
G. de Nittis and K. Gomi, Chiral vector bundles: {A} geometric model for Class {AIII} topological quantum systems, math-ph/1504.04863.

\bibitem{fruchart2014parallel}
M. Fruchart, D. Carpentier, and K. Gaw{\k{e}}dzki, Parallel transport and band theory in crystals, {\it Europhysics Letters} {\bf 106} (2014) 60002.

\bibitem{prodan2014non}
E. Prodan, The Non-Commutative Geometry of the Complex Classes of Topological Insulators, {\it Top. Quant. Matter} {\bf 1} (2014) 1--16.

\bibitem{park2008complex}
E. Park, {\it Complex topological ${K}$-theory} (Cambridge Univ. Press, Cambridge, 2008).

\bibitem{gracia2001elements}
J. M. Gracia-Bond{\'\i}a, J. C. V{\'a}rilly, and H. Figueroa, {\it Elements of noncommutative geometry} (Birkh{\"a}user, Boston, 2001).

\bibitem{prodan2014non2}
E. Prodan and H. Schulz-Baldes, Non-commutative odd {C}hern numbers and topological phases of disordered chiral systems, math-ph/1402.5002.

\bibitem{karoubi1978k}
M. Karoubi, {\it ${K}$-theory: {A}n Introduction} (Springer-Verlag, Berlin, 1978).

\bibitem{blackadar1998k}
B. Blackadar, {\it ${K}$-theory for operator algebras} (Cambridge Univ. Press, Cambridge, 1998).

\bibitem{nicolaescu1997generalized}
L. I. Nicolaescu, {\it Generalized symplectic geometries and the index of families of elliptic problems} (Amer. Math. Soc., Providence, 1997).

\bibitem{higson2000analytic}
N. Higson and J. Roe, {\it Analytic ${K}$-homology} (Oxford Univ. Press, Oxford, 2000).

\bibitem{avron1990quantum}
J. E. Avron, R. Seiler, and B. Simon, Quantum {H}all effect and the relative index for projections, {\it Phys. Rev. Lett.} {\bf 65} (1990) 2185--2188.

\bibitem{avron1994charge}
J. E. Avron, R. Seiler, and B. Simon, Charge deficiency, charge transport and comparison of dimensions, {\it Commun. Math. Phys.} {\bf 159} (1994) 399--422.

\end{thebibliography}
\end{document}